# Extreme exoworlds and the extremophile paradox

Ian von Hegner
Future Foundation Assoc.
Egedal 21
DK-2690 Karlslunde

**Abstract** Extremophiles have gained prominence by providing an experimental approach to astrobiology. Extremophiles gain equal value by being part of a framework for high-level characterisation of the evolutionary mechanisms that must necessarily restrict or promote their emergence and presence on solar system bodies. Thus, extremophiles exist in extreme environments, and therein lies the paradox: extremophiles can only live in extreme environments but yet are not able to originate in such environments. Therefore, even though the range of extremophile capabilities in extreme environments is wider than that in mesophiles, the range of their emergence possibilities is still equally restricted. Therefore, even if one locates an extreme exoworld where terrestrial extremophiles could live here-and-now, it can be predicted that no extremophile analogues are present anyway. Furthermore, it is possible for a world to be uninhabited, yet be habitable, and therein arises the extreme environment paradox: an extreme environment can sustain chemical evolution as well as arriving non-native life, yet native life cannot be built up in that very environment. Thus, life may exist on an extraterrestrial extreme world (if imported there), and chemical evolution may be present on that world. However, it can be predicted that there is no native life anyway. This situation can be predicted to function as a chemosignature and eventually as a biosignature. However, the fact that a non-native extremopile in principle can exist in extreme environments may demonstrate that the intermediate step between chemical evolution and extremophiles can still occur in the form of a statistical deviation. In summary, the use of extremophiles as analogues to extraterrestrial life has limitations due to the very conditions evolution operates under, although analysis of these conditions provides conceptual tools for the search for life elsewhere in the Solar System and beyond.

## 1. Introduction

Astrobiology is a multidisciplinary scientific field that investigates the deterministic conditions and contingent events in which life arises, spreads, and evolves in the universe [von Hegner, 2021c]. The field is wide, with many directions and subjects. One of these subjects includes organisms known as extremophiles. These can, among many definitions, be defined as 'an organism that is tolerant to particular environmental extremes and that has evolved to grow optimally under one or more of these extreme conditions' [von Hegner, 2020a].

Extremophiles have gained prominence not only by providing an experimental approach to astrobiology and valuable data but also by being able to act as analogues to potential extraterrestrial life. This seems reasonable, as extremophiles show that the capabilities and ecological range of life's possibilities are greater than what was assumed before their existence became known.

Thus, among extremophiles that can be mentioned in terms of potential life in extreme worlds is *Deinococcus geothermalis DSM 11300,* capable of surviving at -25 °C [Frösler et al. 2017], or *Planococcus halocryophilus Or1,* capable of growing at -15 °C with 18% salinity [Mykytczuk et al., 2013]. That there are organisms that can obtain such capacities is relevant, e.g., worlds such as Jupiter's moon Europa and Saturn's moon Enceladus both possess liquid suboceans beneath their ice shells, where some regions may have temperatures near the freezing point [Melosh et al., 2004; Sekine et al., 2015].

At the opposite end of the scale can be mentioned extremophiles such as *Geogemma barossii 121,* organisms capable of living at temperatures up to 130 °C [Kashefi and Lovley, 2003]. That there are organisms that can obtain such capacities is relevant as well, as it was estimated that hydrothermal activity at temperatures greater than 90 °C may exist on the seafloor of Enceladus [Sekine et al., 2015].

Mention can also be made of *Chroococcidiopsis sp.*, an organism that can endure several cycles of drying and wetting and is capable of prolonged desiccation in extremely arid hot and cold deserts [Billi et al., 2001], which are relevant when, e.g., looking for life on worlds such as Mars.

While extremophiles have been valuable in providing an experimental approach to astrobiology, they can also be equally valuable in entering into a framework for high-level characterisation of the evolutionary mechanisms at play, where analysis of example scenarios allow us to provide predictions of what will happen in certain situations. Thus, they can provide excellent scenario abstractions and characterisations that tell us what must necessarily restrict or promote extremophiles and thus their presence on solar system bodies such as planets or moons and perhaps even on rogue planets when evolutionary mechanisms and processes are included. These scenario abstractions and characterisations can also tell us what must apply to the emergence or continuation of extremophiles in a world and indeed to the very phenomena that produce them.

Such an approach will be taken in this work, where evolutionary theory will be used as a framework for high-level characterisation of a number of example scenarios. No specific extremophiles will be used in the analyses, but instead an attempt to be generic will be made.

Importantly, when referring to extremophiles, only microbial organisms such as bacteria or archaea are meant here. Thus, the well-known tardigrade will, e.g., not be mentioned here. Although it is often referred to as an extremophile, it is important to bear in mind that it does not have a life cycle in an extreme environment but is merely extremotolerant. Therefore, according to the definition of extremophiles given earlier, only microbial organisms that have evolved to grow optimally under extreme conditions will be considered.

In the following work, a number of realistic constructed scenarios for life and its conditions will be presented, which will provide both a conceptual framework and conceptual tools useful for future focused structured observations and analyses within astrobiology.

**2. Discussion**

In this paper, Section 2.1 presents and clarifies the extremophile paradox; that is, that extremophiles can only live in extreme environments but yet are not able to originate in such environments. Section 2.2 presents and clarifies the extreme environment paradox; that is, that an environment can sustain chemical evolution as well as incoming non-native life, yet native life cannot built up in such environments. Section 2.3 presents and discusses that while an extreme environment severely restricts the seemingly probable events of chemical evolution and biological evolution, an improbable event may still occur, allowing an extremotolerant organism to emerge in an extreme world. Finally, Section 3 summarises the results of this work and its implications for the use of extremophiles in the search for extremophile analogues in the Solar System and beyond.

*2.1. The extremophile paradox*

In astrobiology, one of the interests regarding extremophiles is derived from the observation that they show that the range of life's possibilities is greater than what was assumed before their existence became known. Thus, knowledge of the capabilities and ecological range of extremophiles will be useful in the search for life elsewhere in the Solar System and beyond, as extremophile analogues could hypothetically exist on extreme worlds.

However, extremophiles are found in extreme environments, and therein lies the dilemma. It was pointed out by von Hegner (2020a) that "even though an extremophile can live in an extreme environment here-and-now, its ancestor however could not live in that very same environment in the past, which means that no contemporary extremophiles exist in that environment" [von Hegner, 2020a].

The conditions that apply are that life as a result of physicochemical necessities arises at the simplest conceivable complexity possible since there is no room for variation in the direction of less complexity [Gould, 1996]. The first autonomous cell that emerged was similar to a bacterium peeled down to its most essential constituents, and it began as the simplest rudimentary functional organism possible.

An extremophile is not the simplest conceivable organism possible but is in comparison to the first life a complex organism that has gained a number of adaptations over time to exist in an extreme environment. The first life on this world and hypothetically on other worlds must have required a relaxed environment to emerge, as such rudimentary functional life in comparison with extremophiles, and even later mesophiles, would have been fragile [von Hegner, 2020a]. Thus, there can be 4 variables at play here: environment, mesophile,[1] adaptation and extremophile.

This can rigorously be formulated as follows: If the environment is extreme, then a mesophile will not be able to emerge. If a mesophile cannot emerge, then it will not be able to evolve to become an extremophile. If a mesophile cannot evolve to become an extremophile, then there are no extremophile in that environment. Thus, if the environment is extreme, then there are no extremophile in that environment.

This line of argumentation can be formalized using sentential logic as:

$$P \rightarrow \neg Q, \neg Q \rightarrow \neg R, \neg R \rightarrow \neg S \models P \rightarrow \neg S \qquad (1)$$

where the key is P: the environment is extreme; Q: a mesophile can emerge; R: a mesophile can evolve to become an extremophile; and S: there are extremophiles in that environment.

The above conclusion is seemingly a (biological) paradox, as extremophiles evidently do exist in extreme environments on the Earth. Thus, extremophiles exist in extreme environments on the Earth because their nonextremophilic progenitor originated in a relaxed environment. Therefore, even though the range of extremophile capabilities in extreme environments is wider than that of mesophiles, the range of their emergence possibilities is equally restricted with that of mesophiles.

Life should not be seen as a single isolated entity. Instead, life should be seen as a distribution, as a series of organisms that spread out and gradually become adapted to a local environment. For example, thermophiles and hyperthermophiles living in almost boiling water have not always lived in this environmental condition. Instead, it is a series of organisms that have gradually approached ever more extreme environments over time, while natural selection has selected for the variants that could gradually cope with the gradually increasing warm local environment (Figure 1a).

The situation can be illustrated with a reverse example. An extreme halophile is an organism whose main characteristic feature is a high salinity requirement for growth and metabolism [von Hegner, 2021c].

It is true that while an extreme halophile can live in the environment here-and-now, its ancestors, however, could not, as its capability is obtained through adaptation over time, where generations of its ancestors, step by step, have approached increasingly extreme environments, where they adapted to the increasing salinity to exist in an environment where its ancestors could not. This means that a contemporary extremophile and its past ancestor did not possess the same capabilities.

Thus, if this extremophile is removed from its hypersaline environment and placed in a different environment, e.g., in pure water, then, as everything inside the extremophile requires a high concentration of salts for maintenance, it will in fact burst open due to the osmotic pressure of the new environment [von Hegner, 2021c]. Thus, life gradually adapts to new local environments, and extremophiles like all life require a gradual adaptation over time for escape and subsequent survival, which is a key insight of Darwinian evolution.

Thus, if an extraterrestrial environment overall and persistently has always been extreme, then a transported terrestrial extremophile could live in that environment here-and-now (Figure 1c). However, there are no native extremophiles in that environment anyway, as a mesophile can only arise and live in a relaxed environment (Figure 1b). As a mesophile gradually evolves to become an extremophile through a series of environmental layers with increasing extremity, mesophiles in the past would not have been able to arise and live if the environment has always been uniformly extreme and would consequently not have been able to evolve to become extremophiles.

---

[1] For convenience, in the future I will call the simplest life a mesophile, although mesophiles strictly speaking possess adaptations that make them more complex than the first life.

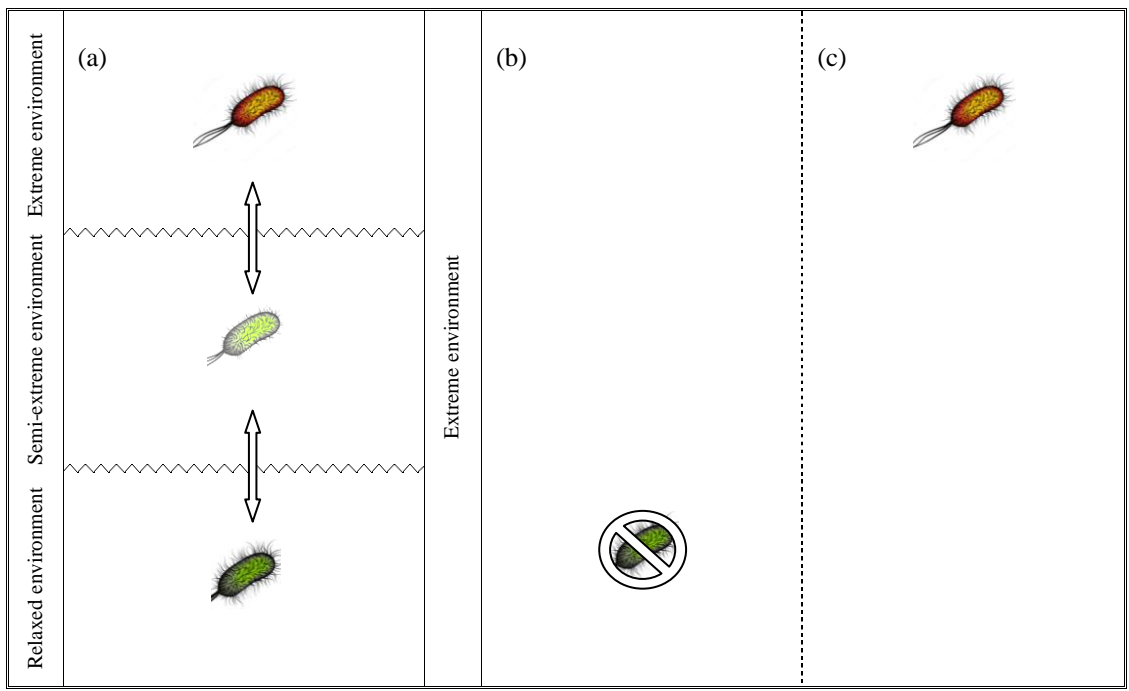

**Figure 1.** The interplay between environmental and evolutionary conditions of life. (a) From bottom to top, the environment is divided into 3 layers with increasing extremity. The different layers illustrate multiple steps of adaptation going from the simplest possible organism to an adapted organism on to an extremophile. The double-headed arrows show that biological evolution does not have a preferred direction but can move from simple to complex and back again. (b) Only an extreme environment exists that is uniform over time and space. Thus, the simplest rudimentary organism cannot arise there, and consequently, no native extremophile can evolve there. (c) Only an extreme environment exists that is uniform over time and space. An imported extremophile from a similar terrestrial environment could potentially survive here, even if its mesophilic ancestors would not have been able to. Credits: bacterial images adapted from Mirumur, 2011, 2014.

This can rigorously be formulated as follows: If an extremophile exists, then it exists in an extreme environment. If an extremophile exists, then it has gradually evolved through a series of environmental layers with increasing extremity from a mesophile. If a mesophile exists, then it can only have emerged and exist in a relaxed environment. Thus, if an extremophile exists in an extreme environment, then it has gradually evolved from a mesophile from a relaxed environment.

This line of argumentation can be formalized using predicate logic as:

$$(\exists x)Ex \rightarrow P, (\forall x)(Ex \rightarrow (\exists y)(My \wedge Axy)), (\exists y)My \rightarrow R \models (\forall x)((Ex \wedge P) \rightarrow (\exists y)(My \wedge Axy) \wedge R) \quad (2)$$

where the key is P: there is an extreme environment; R: there was a relaxed environment; Ex: x is an extremophile; Mx: x is a mesophile; and Axy: x has evolved from y.

Thus, this more adequate analysis than the one given by formalization (1) shows that even though an extremophile can only live in an extreme environment and even though the environment and extremophiles fit together, there are no extremophiles in the extreme environment anyway, as its ancestor, a mesophile, could have only arisen in a relaxed environment and evolved from there. Thus, extremopoiesis emerges from mesogenesis rather than mesopoiesis emerging from extremogenesis.

Therefore, even if one locates an extreme exoworld that meets all the conditions for life as we know it, possesses energy sources, a certain distance to the star, and possesses an environment where one can deduce that terrestrial extremophiles could live here-and-now and extrapolates that extremophile analogues may be present there, then there are no extremophiles anyway due to the very conditions evolution operates under.

Thus, it is, seemly paradoxically, the very extreme environment that prevents extremophiles from occurring in that environment.

*2.2. Uninhabited yet habitable worlds*

An interesting situation arises in continuation of the scenario discussed in the previous section. In this realistic constructed scenario, it was discussed that an extremophile, as seen in Figure 1c, but not a mesopohile, as seen in Figure 1b, may exist in an extreme environment on a solar system body. However, interestingly, if an extremophile in principle can exist in that environment, then chemical evolution may also, even if the step between an extremophile and chemical evolution, a mesophile, cannot (Figure 2b). For our purpose, it is clarified here that by the term extreme environment is meant an environment in which at least one known terrestrial extremophile is capable of living here-and-now.[2]

There have been discussions that it is possible for a world to be uninhabited but yet habitable [Cockell et al., 2012a]. Thus, in clarifying what constitutes an extreme environment, we do not consider worlds that, e.g., are solid frozen throughout, but we keep in mind that it is uninhabited, yet habitable worlds we apply our scenario abstractions and characterisations to.

That a world is uninhabited can be straightforward. However, that a world can be uninhabited and yet at the same time be habitable is a relationship that can be puzzling. If the emergence of life, e.g., is considered to be a deterministic process occurring under the right conditions and these conditions are present on that world, then it is strange that that world is habitable yet uninhabited.

This can be explored through, in principle, a possible test, here called 'the terrestrial test', which predicts that a series of successive situations will be observed.

If, through technological means, one places a terrestrial microbial organism 'a' in an extraterrestrial world's uninhabited environment, then one has subsequently in a trivial sense a binary situation given by a ∈ {0, 1}, which means that a is either 0 or 1, or a ∈ $Z_2$. Thus, either the terrestrial organism will survive or it will not.

However, if it survives in a similar environment as its native environment,[3] then one has a new and nontrivial binary situation, a viability test that allows for two related yet distinct predictions to be made. It can survive for a short time, drawing on limited internal or external resources, or it can continue to live, eventually becoming the progenitor of a clonal colony.

If it continues to live, then this shows that chemistry favourable for life as well as energy sources accessible for life are present in the world from which it consistently is drawing. The latter is important because the fact that the microbial organism can be integrated into the extraterrestrial environment in itself is an indicator that more than chemistry is going on. For rather than saying that a world possesses life, it is more accurate to say that a world processes life. This is an important distinction.

Thus, if the organism merely survives a short time on that world, drawing on internal or external resources, then that world possesses life.

However, if the microbial organism continues to live, eventually becoming the progenitor of a clonal colony, then that world processes life. Life forms do not exist cut off from the environment but are deeply and reciprocally interwoven in it.

If the organism continues to live, this shows that chemistry favourable for life as well as energy sources accessible for life are present as it draws from these resources. Chemistry and chemical evolution are not the same; chemical evolution presupposes chemistry, but chemistry does not presuppose chemical evolution.

---

[2] It may be stated that one cannot say in advance that there are no extremophile analogues that can withstand environments that terrestrial extremophiles cannot cope with, nor that terrestrial extremophiles will not be discovered in the future that push the boundaries of life even further. However, while being valid points, the evolutionary conditions discussed here will still remain in place.

[3] Despite the fact that extremophiles are considered generally hardy organisms, from an evolutionary perspective they are only locally hardy. As for mesophiles, they are adapted to their specific local environment. Thus, there is an inverse proportionality between two environments at stake here in the form that the environment in one world must match the environment in another world [von Hegner, 2020c].

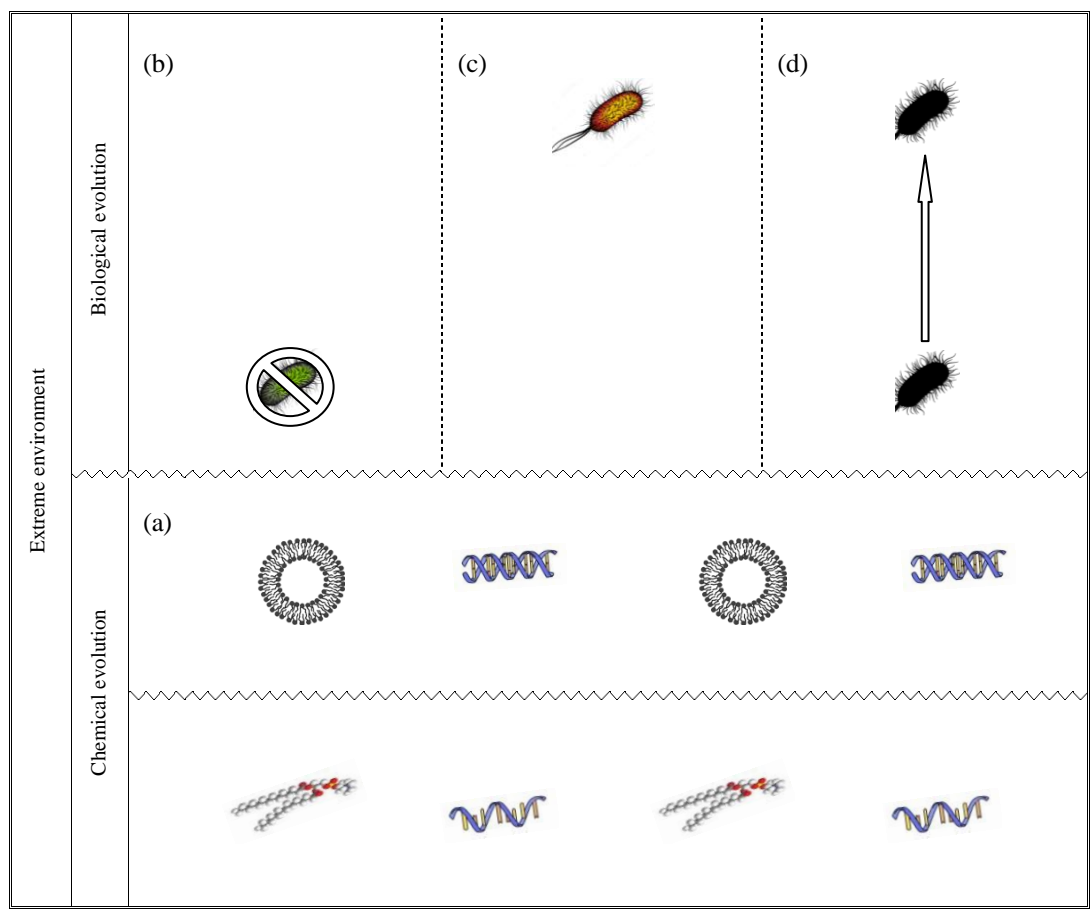

**Figure 2.** The steps of chemical evolution that proceed the steps of biological evolution. (a) A simplified view of the series of quasi-biotic layers leading to the production of the first rudimentary functional cell (phospholipids, RNA strands, liposome cell membrane and DNA molecule shown are not on a mutually correct scale). (b) A simple rudimentary organism cannot arise in an extreme environment, and consequently, no native extremophile can evolve there. (c) An imported extremophile living in an extraterrestrial extreme environment even though its ancestors would not be able to live there. (d) The emergence of a 'hopeful monster,' an extremotolerant organism that can withstand the environment in a single improbable leap. Credits: bacterial images adapted from Mirumur, 2011, 2014.

However, if chemistry favourable for life as well as energy sources accessible for life are present, it seems reasonable to assume that the conditions for chemical evolution are present as well.

Thus, the following conjecture can be proposed: if chemistry favourable for life as well as energy sources accessible for life are present in a habitable world, then the conditions for chemical evolution are present as well.

That chemical evolution cannot produce an autonomous cell in a habitable world is a nontrivial statement, as if a transported organism can survive there and indeed establish itself and eventually become the progenitor of an entire biota in the hitherto uninhabited world, then it is not obvious why native life did not arise there.

There may be suggested several reasons why there is no life in such a world. However, the focus here will only be on evolutionary reasons based on an arriving organism. It is true that both mesophiles or extremophiles can be placed in an uninhabited yet habitable world, and their initial survival depends on whether it is a relaxed or extreme environment similar to their own environment in question.

If it is a relaxed world and a transported terrestrial mesophile can live there and become the progenitor of a biota, showing the possibility of biological evolution, then the conditions for chemical evolution are present. The fact that native life has not arisen may indicate that chemical evolution is not a fully deterministic process but includes contingent events as well.[4] This is the case for biological evolution, and there appears to be no reason to assume the same does not influence chemical evolution. Thus, although many of the steps of chemical evolution may be solely physicochemically driven – they will inevitably happen when the conditions are present – there are other steps that may depend on contingent events.

If it is an extreme world and a transported terrestrial extremophile can live there and become the progenitor of a biota, showing the possibility of biological evolution, then the conditions for chemical evolution are present. The fact that native life has not arisen can indicate that chemical evolution, in addition to not being a fully deterministic process, cannot produce an autonomous cell under extreme conditions, which is what will be focused on here.

The first life is, as mentioned in the previous section, the simplest functional organism possible; it is a fragile organism as opposed to its descendants who have evolved more robust traits. Thus, extremopoiesis emerges from mesogenesis.

Thus, although an extremophile in principle can exist in that environment and chemical evolution in principle can also occur there, the environment is too extreme for the necessary step between chemical evolution and an extremophile, or a mesophile, to occur there. The step between a mesophile and an extremophile requires a series of adaptations, as shown in Figure 1a. However, the step between a mesophile and chemical evolution also requires a series in which the assembly of cellular components occurs, as shown in Figure 2. If the environment has always been extreme, then chemical evolution will not be able to complete the assembly of a native cell and consequently no extremophiles. Therefore, even though the steps of chemical evolution can be solely physicochemically driven, some of these steps cannot occur under extreme conditions. Thus, some of the components of a cell, such as RNA or DNA, cannot occur in an extreme environment.

Thus, analysis of this example scenario shows, seemly paradoxically, that life may exist on an extreme solar system body (if imported there) and that chemical evolution may be present on that world. Yet there is no (native) life there anyway, as the extreme environment itself prevents it, and this despite the fact that the extreme environment is conducive to both extremophilic life forms and chemical evolution.

*2.2.1. Chemosignatures*

The situation discussed in the previous section, which is based on the very conditions of how evolution works, can be predicted to function as a chemosignature. The situation illustrated in Figure 1a. does not surprisingly lead to a biosignature. However, the situation illustrated in Figure 2a,c. can more subtly lead to a chemosignature and eventually to a biosignature. A chemosignature is almost analogous to a biosignature, where one does not look for signs of life in a world but for signs of whether chemical evolution can occur on a world.[5]

However, there will be a connection between a chemosignature and a biosignature that can in principle be tested. The conjecture had previously been granted that if chemistry favourable for life as well as energy sources accessible for life is present in a world, then the conditions for chemical evolution are present as well.

On such an exoworld, an imported organism that thrives eventually will make its presence noticed because one of the most distinctive features of life is that it goes 'information explosive' [von Hegner, 2021c]. As soon as life arises in a world, it will undergo exponential growth, increase structured information, and gradually fill all available niches in a world.

Thus, if a microbial organism arrives in an uninhabited but habitable world, it too goes 'information explosive' there and can in principle fill that world up through its exponential growth relatively quickly. If we take the content of water on the moon Europa as an example. The liquid ocean underneath the moon's ice shell may be approximately 100 km deep [Chyba and Phillips, 2002], with a volume estimate as high as $3 \times 10^{18}$ m$^3$.

---

[4] Life can also occur but be prevented from continuing due to the indeterminacy bottleneck [von Hegner, 2021c].

[5] Just as it is the case with many suggested biosignatures, where it can indicate the presence of life, but it cannot be ruled out that other abiotic factors produced them, so it can also be the case here that it is only chemistry and not conditions for chemical evolution that are present.

Thus, how many extremophiles can such a world contain?

The size of organisms, such as bacteria, ranges between ~0.4 and 3 $\mu m^3$ [Levin and Angert, 2015]. Thus, it can be estimated for convenience that the extremophile organism has a length of 1.5 μm and a diameter of 0.5 μm. Estimating the volume of the extremophile using a cylinder, we find the volume of a single extremophile is: $V_2 = \pi r^2 l = 2.94 \times 10^{-19}$ m$^3$.

We already have the estimate for Europa's oceans: $V_1 = 3 \times 10^{18}$ m$^3$. Thus, the volume of the moon's ocean is divided by the volume of a single extremophile:

$$\frac{V_1}{V_2} \quad (3)$$

which shows that $1.02 \times 10^{37}$ extremophiles could fit in the moon's ocean. However, the extremophiles are not perfectly formed structures that are neatly packed next to each other. Thus, we estimate that the moon's ocean as a result of the gaps between the extremophiles contains 15% free fluid space by volume. We use proportions to calculate how many extremophiles would take up 85% of the ocean by volume:

$$\frac{x}{1.02 \times 10^{37} \text{ extremophiles}} = \frac{85\%}{100\%} \quad (4)$$

$$x = 8.67 \times 10^{36} \text{ extremophiles.}$$

The extremophile reproduces by binary fission, where its numbers increase exponentially, given by:

$$N_n = N_0 2^n \quad (5)$$

where $N_n$ represents the number of extremophiles at the end of the time interval, $N_0$ represents the initial number of extremophiles at the beginning of a time interval, and n represents the number of generations. Applying this formula shows that it will only take 122 cell divisions to fill up that world's ocean with $N_n = 10^{36}$ descendants of a single arriving extremophile.

In a world with an already existing ecosystem, the number will obviously not be so high. However, in the example here, the world is habitable but uninhabited; thus, there are initially no competitors, plenty of available space, free energy and chemical building materials. The descendants of one organism will become competitors, but if there is so much space and free resources available, it may come a long way in filling up that world's ocean before it becomes an issue. Thus, here, a chemosignature will relatively quickly be able to demonstrate a biosignature, as life not only exists passively on a world but also affects it and thus that world is conducive to chemical evolution.

A more exotic biosignature can also exist, as life will not necessarily be restrained to that world afterwards, as life in certain situations, so to speak, 'leaks' from the world. On solar system bodies such as the Earth, an external violent event is required for life to obtain escape velocity from its world, which can occur in the form of lithopanspermia, or an internal violent event such as space craft-mediated transport [von Hegner, 2021a].

However, in worlds such as Enceladus, which have been observed to emit jets of mainly water from its south polar region [Porco et al., 2006], potential organisms living in its ocean are likely to be occasionally sprayed up from the plumes [Porco et al., 2017] and thus 'pollinate' the space with biomatter. Due to the low gravity and water jets of such worlds, they may indeed be feeding the nearby space with organisms. Thus, Saturnian space may in principle contain satellite biobanks, which may also be embedded in the planet's E ring; thus arriving or leaving, due to life's distinctive feature that it goes 'information explosive', it will make its presence noticed.

*2.3. Variants*

If the environment on a solar system body has always been extreme, then extremophiles, as discussed in the scenarios in the previous sections, would not have been able to occur there, as their nonextremophilic ancestors could not arise and live in that environment and thus evolve to become extremophiles.

However, the fact that a non-native extremophile in principle can exist in the extreme environment, demonstrating that the conditions for biological evolution and chemical evolution are present, even though the intermediate step between them, a mesophile, is not there, raises an interesting question: Is there necessarily an absolute conflict between the simplest organism possible and an extreme environment? Can this intermediate vital step between chemical evolution and extremophiles with the assembly of a cell that can resist the environment and have time to adapt to it, in principle still be possible as an exception to the rule, as a rather modest statistical deviation, thus satisfying the principle of plenitude.

It is, *prima facie,* not easy to see how to construct a realistic scenario for how the simplest life could arise in an extreme environment, but there are two things that seem to apply in that regard.

First, as seen in Figure 1a, biological evolution proceeds through a series of adaptations, from the simple rudimentary functional cell to an evolved organism and from there on to an extremophile. This occurred in a series of layers with increasing extremity or environmental stressors. However, it is also the case that before the emergence of the first functional cell, before biological evolution begins in this first layer in Figure 1a, there are a number of layers with quasi-biotic steps in which the assembly of a cell through several layers occurs; namely, chemical evolution, as seen in Figure 2.

Second, the first autonomous cell is not a platonic absolute but is a variant. Biological evolution addresses individual variation and natural selection among organisms. However, variation does not necessarily begin with the first rudimentarily functional cell. The progenote designates precells preceding the cenancestor [Gogarten, 2019], which means there may also have been variants among the precells that came before the first functional cell. Thus, the first autonomous cell in one world will be the simplest possible organism, but it will still be able to exhibit modest variation compared with the first autonomous cell in another world. They will be almost identical but with modest space for variation between them. Thus, extending a Darwinian principle, where no specific species occupies a privileged position in nature, there is also nothing privileged regarding even the first autonomous cell on the Earth.

With these two abstractions in mind, a series of layers in which quasi-biotic steps occur and the first autonomous cell occurs as a variant, it could be imagined that rather than as in the previous scenario where evolution occurred from mesophiles to extremophiles, evolution can also occur from extremotolerant organisms to extremophiles.

Thus, it can be hypothesised that a significant but possible deviation is possible where a cell on the boundary between progenote and autonomous cells achieves extremotolerant capacities that barely allow it to survive the extreme environment (Figure 2d). While mesophiles and extremophiles have their optimal growth conditions in their local environments, an extremotolerant organism differs from them in being able to tolerate an environment but not thrive in it.[6]

In this specific scenario, such an event could be considered a case of a saltational event, where the improbable extremotolerant or even polyextremotolerant organism is in some ways analogous to a so-called 'hopeful monster', an organism that makes a single leap without intermediate adaptations and selection and happens to be viable in the specific environment. Saltational events have been discussed for a long time, mostly unfavourably, although there is support for individual cases [Katsnelson et al., 2019].

There is still much to know regarding chemical evolution, and therefore, no attempt will be made to dwell on the exact chemical steps here. However, the demand that the cell must arise in a relaxed environment is not accidental. Many of the subprocesses of chemical evolution are delicate and will be destroyed in extreme environments. Thus, many of the structures that are part of a cell, such as RNA and DNA, are relatively fragile.

However, in the following scenario, it will be assumed for the sake of argument that in many of the layers of the environment and in many of the phases of chemical evolution, chemical evolution can occur under extreme conditions and that it is only in the last layer, the last phase, that the cell is destroyed. This is based on, as discussed in the previous section, that if chemical evolution can be present on a solar system body, then the possibility of its phases must necessarily be present as well.

---

[6] Life attains extremotolerant capacities before they attain extremophilic capacities, adapting to a new environment step by step as they slowly approach the extreme environment. Thus, if such an extremotolerant organism emerges and remains active in the new environment, then its descendants will also eventually evolve to become extremophiles.

Environmental stressors destroy the simplest functional cell that arises as an end product of chemical evolution, although they do not necessarily destroy the steps that came before the last layer. Thus, it will be assumed here that many progenote variants can occur simultaneously in an extreme environment, and rather than all variants being destroyed in the last layer, there could in principle be a single variant that achieves extremotolerance.

It is also pointed out here that the extreme environment in the last layer is not a uniform quantity. The term extreme environment requires a clarification of the type of environment in question. Thus, there are many different types of extreme environments, such as cold, hot, low salinity, high salinity, etc. Some of these different types of environments can also come in combinations of each other with varying ratios between them.

There is no universal extremophile, as these different combinations result in a variety of extremophiles needed to cope with any given environment. However, this has no bearing on the discussion in the previous sections, as the evolutionary conditions discussed here remain in place. Thus, the environment is divided into different components, and it is assumed here that chemical evolution can occur with certain combinations of these components present. Specific components are not discussed here; they are only discussed here in a generic sense.

These two assumptions are a simplification but are here assumed for the sake of calculation.

Thus, that a cell on the border between progenote and autonomous cells achieves extremotolerance that allows it to survive the environment depends on two things:

That in the last layer, the cell obtains a certain ordered combination of some components from its own pool of different components, where the pool for the sake of convenience is n = 56.

That the extreme environment in the last layer is not a uniform quantity but is divided into different components, each of which are stressors on life or its origin, which here for the sake of convenience is k = 6.

Thus, there must be an alignment of interest between exactly the specific combination of the components of the cell and the specific components of the extreme environment. In evolutionary theory, the alignment of interest is usually between the organism's internal components of joint reproduction or between the organisms. However, it could in principle also occur among the components of the cell and environment, showing that this does not necessarily have to be considered in terms of random chance.

Thus, the cell's combination of components is similar to a lock with its set of obstructions or wards, and the environment is similar to a key with its notches or slots corresponding to the obstructions in this particular lock, where the cell's combination matches the individual components of the extreme environment. A match means that the cell, the lock, with its wards with bends and complex protrusions can withstand the environment, as the notches or slots in the key correspond to the wards in the lock and the key will therefore turn, allowing the cell to survive and even reproduce even if it is not an extremophile.

In other words, in this deviant scenario, in its final phase, the cell must achieve a random composition, a precise combination that must be a fit with the surrounding environment to proceed further. This will not be a preadaptation but more in line with an exaptation, as it cannot be ascribed to the direct action of natural selection, as the cell simply happens to be almost improbably fine tuned to have a match to exactly the environmental conditions. Abiogenesis does not occur in a single linear leap, but that is not what is discussed here either. Here, it is assumed that a large group of progenotes already existed simultaneously in the last layer.

Thus, what is the probability that chemical evolution in a single leap can produce an extremotolerant variant of a progenote with a combination that exactly matches the extreme environment?

For one progenote variant to obtain extremotolerance, it has to match six numbers or components from the environment with numbers from its own pool of 56 numbers or components ranging from 1 to 56. To calculate the probability for 6 matching components, we have for the entire situation, that n = 56, the total number of possible choices, k = 6, the number of choices that can be made. There are more components in the cell than in the environment. Therefore, the cell is here considered symbolically as the lock, and the environment as the key. To find the total numbers of possible combinations for any 6-digit number, we use the formula for combination in the following:

$$C(n, k) = \binom{n}{k} = \frac{n!}{k!(n-k)!} \tag{6}$$

$$= \frac{56!}{6!(56-6)!}$$

$$= 3246843601.$$

Thus, ~32 million are the odds or the total number of possible combinations for the 6-digit number to obtain extremotolerance, where the probability found by just dividing 1 by the number above is $3.07 \times 10^{-8}$.

In that calculation, order was not included. However, when it is important as is here, permutation should be used if the components in the environment must match the order of the string of components granting extremotolerance in the cell. Thus, we use the permutation formula in the following:

$$P(n, k) = \frac{n!}{(n-k)!} \tag{7}$$

$$= \frac{56!}{(56-6)!}$$

$$= 2.337727392 \times 10^{10}.$$

Thus, ~23 billion is the odds of getting the combination that gives extremotolerance, where the probability of obtaining extremotolerance for this case, which is found at 1 divided by the odds, is $4.27 \times 10^{-11}$.

Calculation (6) shows that the probability of obtaining the right match in an extreme environment is 1 in 32 million. Thus, in a galaxy such as the Milky Way that is estimated to contain 100-400 billion stars [Williams, 2008] and hence even more planets and moons, the possibility could in principle occur. Even the more accurate calculation (7), where the probability of obtaining the right match in an extreme environment is 1 in 23 billion, could also in principle occur in the galaxy, depending on how many of these planets and moons are habitable.

Thus, if the constructed scenario above is valid, then it shows that chemical evolution has more than one approach to the origin of life available: an improbable one and, in comparison, a probable one.

However, next to the fact that saltational events with drastic single leaps are debated, the fact that the probabilities were so low with so few components used, we assumed only n = 56, plays a role, as a more realistic constructed scenario with considerably more components as a cell contains would give a much lower probability, probably demanding more habitable worlds contained even in an entire galaxy group such as the Local Group to occur, effectively ruling such hopeful monsters out of the picture.

**3. Conclusion**

Astrobiology is not merely the search for life elsewhere in the Solar System and beyond (although the discovery of life elsewhere would of course be immensely important). It is instead fundamentally the study of the conditions that underlie the origin of life and evolution in the universe. Thus, the underlying study objectives of astrobiology are fundamentally chemical evolution and biological evolution and the conditions with which they occur [von Hegner, 2021c].

One objective for it is, e.g., to seek to distinguish between universal necessary features of all life and features that have merely arisen due to historical circumstances; that is, by contingent events. To date, only one example of life is known, namely, terrestrial life. There is therefore a justified discussion regarding which features are universal and which, due to specific circumstances of the Earth, apply to life.

However, the extremophile paradox shows an evolutionary mechanism that must be universally applicable, thus allowing us to place some restrictions on our understanding and search criteria for life elsewhere in the galaxy.

Thus, the use of extremophiles as analogues to extraterrestrial life has, despite their increasing importance in astrobiology, a limitation due to the very conditions evolution operates under. Therefore, searching for extreme worlds that meet all the conditions of extremophilic life in terms of energy sources, a certain distance to the star, an environment favourable for certain terrestrial extremophiles, etc., is only secondary here. On the basis of evolutionary principles, we can deduce that if a solar system body had always been unfavourable for life overall, then life would not have been able to evolve over time, and extremophiles would not have emerged. The primary criterion is thus to distinguish between worlds possessing local relaxed or extreme environments and worlds that have always been uniformly extreme.

Thus, even if abiogenesis was to be a fully deterministic process, a natural consequence of star formation that is automatically initiated when the conditions are present in a set of specific worlds, then no life would be present in such worlds anyway, precisely because of the extreme environment. These are the seemingly counterintuitive predictions that can be made on the basis of evolutionary theory, although it is not a true paradox when properly analysed.

Another objective of astrobiology is that while clarifying whether a solar system body possesses life is immensely important, from a purely scientific point of view, it is equally important to examine whether a world possesses conditions for life.

The placement of a terrestrial extremophile on an uninhabited, yet habitable world obviously does not represent a search for life *per se*, since one has planted this life there. However, it is a search for conditions of life. This constructed scenario allows us to analyse and limit the restrictions that apply to chemical evolution and biological evolution and thus refine our understanding of the origin of life and its evolution.

The objective was not on how life was transported there; the focus was on the case that an extremophile could live there while a mesophile could not. However, if one considers it an issue with an imported organism, then this step can in principle be naturally simplified by eliminating one of the worlds and the imported organism. In the earlier stages of a world's history, it may be subjected to heavy bombardment resulting from the formation of the Solar System. Thus, even if a world is sufficiently favourable for life to arise early on it, then it can still be hit by a global impactor that can completely sterilize that world's environment. However, by such impacts, matter can be sent up and after a few thousand years return to the world again [Wells et al., 2003].

It has been hypothesised that life during such an event may be sent up inside such material, and after some time in this 'refugium' return and reseed the world it came from [Sleep and Zahnle, 1998]. Life can even achieve a boost in its capacities due to this event [von Hegner, 2021b]. Thus, a solar system body may initially have obtained areas with relaxed environments in which life has arisen that evolve into extremophiles in the adjacent more extreme environments, after which some of these have been sent up by an impactor and later return to a changed world that now has a uniform extreme environment.

Interestingly, if that world initially meets all the conditions for chemical evolution but experiences being reshaped by a global impactor, will that world, once it has stabilised again after millennia, again experience that the conditions for chemical evolution will resurface? Thus, are the conditions for chemical evolution in an abstract sense present all the time, or must chemical evolution itself evolve on a world, meaning that probability also applies to it, even on a habitable world? Thus, more generally, if chemical evolution itself is subject to evolvability, as has been suggested for biological evolution, then this is possibly a further reason that may be proposed for what was asked in the section on uninhabited, yet habitable worlds, namely, that if an arriving organism can survive on that world, then it is not obvious why native life did not arise there.

The situation illustrated in Figure 1a. does not surprisingly lead to the existence of a biosignature in a world. However, the situation illustrated in Figure 2a,c. leads more subtly to a chemosignature and eventually to a biosignature, which provides conceptual tools that allow us to clarify on a deeper level what is meant by habitable worlds.

If a world's environment is extreme, appropriate for certain terrestrial extremophiles here-and-now but has always been uniformly extreme, then native life will not exist there, even if the conditions for chemical evolution are present. Thus, although there is still much to know about the possibilities of biological evolution and the mechanisms of chemical evolution, this allows us to differentiate between habitable environments where chemical evolution occurs and thus refine our search parameters for habitable worlds elsewhere by placing some restraints on these parameters.

As mentioned in the section variants, the emergence of an extremotolerant organism is not to be regarded as a preadaptation but more an exaptation, as it simply happens to be almost improbably fine tuned to have a match to exactly the environmental conditions. Thus, it is important here to point out that abiogenesis does not occur by a cell spontaneously arising fully completed. Instead, the emergence of the first autonomous cell did not occur in merely sequential trials but rather in simultaneous trials.

Thus, in a relaxed environment such as on the Earth, there probably existed a large group of precells, progenotes, with an exchange of genes between them probably occurring, and where natural selection took place, of which a progenote eventually became a fully autonomous cell, after which it or one of its descendants became the progenitor of all later life. Probabilities are at stake here, but presumably not staggering improbabilities.

Indeed, life arose relatively quickly on the Earth, possibly as early as 4.1 billion years ago, in the Hadean [Bell et al., 2015], which is interesting since the Earth originated 4.5 billion years ago, where the majority of the Earth must have been an extreme environment. However, following the Copernican principle, there is no reason to assume that there was anything privileged or unique about the Earth at that time. Thus, life may have originated in spots with relaxed areas where the various phases of life could occur, a so-called 'Great prebiotic spot' that a planet or moon may hypothetically possess [von Hegner, 2020b].

Abiogenesis does not occur in a single linear leap, but that is not what was discussed in the section on variants. Here, it was assumed that a large group of progenotes already existed simultaneously in the last layer. In a relaxed environment, the issue is not an extreme environment. The progenotes have survived thus far because the environment is favourable to them. However, in an extreme world, this implies that many cells may have emerged almost simultaneously and disappeared again, as they could not cope with the environment. However, one of these will then, through a statistical deviation, achieve a match that allows it to survive the environment and to thus continue its evolution and become an autonomously reproducing cell as if this had taken place in a relaxed world.

The calculations in the section about variants show that the origin of life this way in an extreme environment does not have a high probability. This scenario depends on the specific assumptions made, and the calculations are simplistic, hampered by the fact that much is not yet known about chemical evolution. That it is only in the last layer that the cell is destroyed was also a simplification. Thus, they will be subject to future empirical and theoretical research. However, they serve as an illustration of the situation, as the purpose here is not to seek to give an accurate estimate but to clarify how life could arise in an extreme world, which is not easy to do. It shows that the possibility may be present but that it is so low that the scenario of relaxed worlds for all practical purposes is in effect.

However, it is interesting to note that this specific scenario in an extreme environment represents a deviation from the standard scenario. This means that the emergence of an autonomous cell seen from this perspective is a probable event if chemical evolution occurs in a relaxed environment. Because in the standard scenario, the precell does not have to make a leap in the last phase to survive, because here it does not have to be a fit in an extreme environment.

Thus, chemical evolution has already produced variant precells that can exist in the environment, natural selection will occur between them, and a functioning autonomous cell will be able to arise in the relaxed environment. In this scenario, the emergence of life in a relaxed environment is thus not necessarily an improbable event, while the emergence of life in an extreme environment is.

## References

Bell, Elizabeth A.; Boehnike, Patrick; Harrison, T. Mark; Mao, Wendy L. (2015). *"Potentially biogenic carbon preserved in a 4.1 billion-year-old zircon"*. Proc. Natl. Acad. Sci. U.S.A. 112 (47): 14518–14521.
Billi, D, Friedmann, EI, Helm, RF, Potts, M: *Gene transfer to the desiccation-tolerant cyanobacterium Chroococcidiopsis.* J. Bacteriol. 2001 Apr;183(7):2298-305.

Chyba, Christopher F., Phillips, Cynthia B.: *Europa as an Abode of Life* Origins of Life and Evolution of Biospheres 32(1):47-68 March 2002.

Cockell, C.S., Balme, M., Bridges, J.C., Davila, A., and Schwenzer, S.P. (2012a) Uninhabited habitats on Mars.Icarus217:184–193.

Frösler, J., Panitz, C., Wingender, J., Flemming, H.-C., and Rettberg, P. (2017). Survival of Deinococcus geothermalis in Biofilms under Desiccation and Simulated Space and Martian Conditions. Astrobiology 17, 431–447.

Gogarten, Johann Peter: The progenote, LUCA, and the root of the cellular tree of life, Handbook of Astrobiology, edited by Vera Kolb, CRC Press, 2019.

Gould, Stephen Jay: *Full House - The spread of excellence from Plato to Darwin,* Harmony Books New York, 1996.

Kashefi, K., and Lovley, D. R. (2003). Extending the upper temperature limit for life. Science 301:934.

La Scola B., Audic S., Robert C., Jungang L., de Lamballerie X., Drancourt M., Birtles R., Claverie J.M., Raoult D. A giant virus in amoebae. Science. 2003; 299:2033.

Katsnelson, Mikhail I. Wolf, Yuri I. and Koonin, Eugene V. On the feasibility of saltational evolution, PNAS October 15, 2019 116 (42) 21068-21075; first published September 30, 2019.

Levin, Petra A. and Angert, Esther R. Small but Mighty: Cell Size and Bacteria, Cold Spring Harb Perspect Biol. 2015 Jul; 7(7): a019216.

Melosh, H.J. Ekholm, A.G. Showman, A.P. and Lorenz, R.D. The temperature of Europa's subsurface water ocean, Icarus 168 (2004) 498–502.

Mirumur, 2011: https://depositphotos.com/vector-images/bacterium.html

Mirumur, 2014: https://depositphotos.com/49770579/stock-illustration-e-coli-bacteria-isolated-red.html

Mykytczuk, N. C. S., Foote, S. J., Omelon, C. R., Southam, G., Greer, C. W., and Whyte, L. G. (2013). Bacterial growth at -15 °C; molecular insights from the permafrost bacterium Planococcus halocryophilus Or1. ISME J. 7, 1211–1226.

Porco, C.C., P. Helfenstein, P.C. Thomas, A.P. Ingersoll, J. Wisdom, R. West, G. Neukum, T. Denk, R. Wagner, T. Roatsch, S. Kieffer, E. Turtle, A. McEwen, T.V. Johnson, J. Rathbun, J. Veverka, D. Wilson, J. Perry, J. Spitale, A. Brahic, J.A. Burns, A.D. Del Genio, L. Dones, C.D. Murray, and S. Squyres, 2006: Cassini observes the active south pole of Enceladus. *Science*, 311, 1393-1401.

Porco C.C., Dones L., and Mitchell C. (2017) Could it be snowing microbes on Enceladus? Assessing conditions in its plume and implications for future missions. *Astrobiology* 17:876–901.

Sekine, Y., Shibuya, T., Postberg, F., Hsu, H. W., Suzuki, K., Masaki, Y., et al. (2015). High-temperature water–rock interactions and hydrothermal environments in the chondrite-like core of Enceladus. *Nat. Commun.* 6:8604.

Sleep, Norman H., Zahnle, Kevin. Refugia from asteroid impacts on early Mars and the early Earth, Journal of Geophysical Research, Vol. 103, No. El2, Pages 28,529-28,544, 1998.

von Hegner, Ian: *Extremophiles: a special or general case in the search for extra-terrestrial life,* Extremophiles (2020a) 24:167–175; Published online 9 November 2019.

von Hegner, Ian: *A limbus mundi elucidation of habitability: the Goldilocks Edge,* International Journal of Astrobiology (2020b), Vol. 19: published online 22 May 2020; preprint published online in arXiv.org 15 October 2019.

von Hegner, Ian: *Interplanetary transmissions of life in an evolutionary context,* International Journal of Astrobiology, (2020c); published online 27 May 2020.

von Hegner, Ian: *Evolutionary processes transpiring in the stages of lithopanspermia,* Acta Biotheoretica, 10 April 2021a; preprint published online in HAL archives-ouvertes.fr | CCSD, 21 April 2020.

von Hegner, Ian: *A trampoline effect occurring in the stages of planetary reseeding,* Biosystems, Vol. 205, July 2021b; preprint published online in arXiv.org 19 May 2020.

von Hegner, Ian: *The indeterminacy bottleneck: Implications for habitable worlds,* HAL archives-ouvertes.fr | CCSD, 10 April 2021c.

Wells, L.E., Armstrong, J.C., and Gonzales, G. (2003) Self-reseeding of early Earth by impacts of returning ejecta during the Late Heavy Bombardment. *Icarus* 162, 38–46.

Williams, Matt: How Many Stars are There in the Milky Way? Universe today, 2008.